# Galvanic Replacement Reaction to prepare nanoporous Aluminum for UV plasmonics

**Denis Garoli[1*], Giorgia Giovannini[2], Sandro Cattarin[3], Paolo Ponzellini[1], Remo Proietti Zaccaria[1,4], Andrea Schirato[5], Francesco D'Amico[6], Maria Pachetti[6,7], Wei Yang[8], Hai-Jun Jin[8], Roman Krahne[1], and Alessandro Alabastri[5]**

1. Istituto Italiano di Tecnologia, via Morego 30, I-16163, Genova, Italy;
2. EMPA Federal Swiss research Institute, St. Gallen, 9014, CH;
3. ICMATE - CNR, Corso Stati Uniti 4, 35127 Padova, Italy;
4. Cixi Institute of Biomedical Engineering, Ningbo Institute of Industrial Technology, Chinese Academy of Sciences, 1219 Zhongguan West Road, Ningbo 315201 P.R. China.
5. Electrical and Computer Engineering, Rice University, 6100 Main Street MS-378, Houston, TX 77005
6. Elettra Sincrotrone Trieste S.C.p.A., S.S. 14 km 163,5 in Area Science Park, 34149, Basovizza (TS), Italy
7. Department of Physics, University of Trieste, Via Alfonso Valerio 2, 34127 Trieste, Italy.
8. Shenyang National Laboraory for Materials Science, Institute of Metal Research, Chinese Academy of Sciences, 72 Wenhua Road, Shenyang 110016 P.R. China
* Correspondence: denis.garoli@iit.it;

**Abstract:** Plasmonics applications have been extending into the ultraviolet region of the electromagnetic spectrum. Unfortunately the commonly used noble metals have intrinsic optical properties that limit their use above 350 nm. Aluminum is probably the most suitable material for UV plasmonics and in this work we show that nanoporous aluminum can be prepared starting from an alloy of $Mg_3Al_2$. The porous metal is obtained by means of a galvanic replacement reaction. Such a nanoporous metal can be exploited to achieve a plasmonic material for enhanced UV Raman spectroscopy and fluorescence. Thanks to the large surface to volume ratio this material represents a powerful platform for promoting interaction between plasmonic substrates and molecules in the UV.

## 1. Introduction

Collective oscillation of electron waves at the metal / dielectric interface, also known as Surface Plasmons, finds extensively applications in different fields such as surface-enhanced Raman spectroscopy (SERS), metal-enhanced fluorescence (MEF), photocatalysis, etc.[1] Plasmonics and its application have been mainly focused to Visible and Infrared spectral regions because of the optical properties of noble metals (Au and Ag) that are typically used. In fact, interband transitions introduce a dissipative channel for Au and Ag plasmon resonances at wavelengths shorter than 550 and 350 nm, respectively. On the contrary, during the last decade there has been an increasing interest in extending plasmonics effects down to UV and deep-UV (DUV) wavelengths[2,3]. UV and DUV excitations can be utilized to perform Raman spectroscopy on biomolecules that have small Raman cross sections in the visible and NIR regions[4],[5]. Additionally, important biomolecules (such as some amino-acids) have intrinsic fluorescence in the UV region[6] and a platform able to enhance the detection limit thanks to plasmonic effects is highly appealing in biosensing[7]. Several alternative materials have been investigated during the recent years; among them magnesium, gallium, indium, ruthenium, and aluminum (Al)[8–12]. Al in particular has been subject of extensive researches and demonstrated to be a promising plasmonic material in the UV and DUV regions[13]. In order to use Al to generate localized surface plasmonic resonances (LSPR) in the UV spectral region, small metallic features/nanostructures have been realized. Such structures can be designed with the help of electron beam lithography (EBL) and focused ion beam (FIB) lithography or achieved with different methods of synthesis of nanoparticles[14],[15]. An alternative approach to generate LSPR is based on porous metal films that, during the last decade, have attracted increasing interest due to their unique very high specific surface area. In a recent work, we reported on the preparation of Mg/Al nanoporous film based on a $Mg_xAl_{1-x}$ alloy[5]. Chemical dealloying is the typical procedure to prepare nanoporous films[16] and in our previous experiments we showed that a phase selective dealloying can be performed in acetic acid in methanol. Unfortunately, the high reactivity to oxidation of Al and Mg does not enable to prepare porous metals with oxygen contents below 14 %. Moreover, it has not been possible to achieve porous metals in pure Al phase. However, H-J. Jin et al.[17], and more recently J. S. Corsi et al.[18] demonstrated that it is possible to prepare porous Al structures by means of galvanic replacement reaction (GRR) starting from an alloy of $Mg_3Al_2$. Electrochemical or chemical synthesis of nanoporous Al in aqueous electrolytes is challenging because Al is very reactive and nanoscale Al ligaments may be rapidly fully oxidized. GRR involves a dissolution of sacrificial metals via galvanic oxidation and concurrent precipitation of second metals onto the sacrificial one via galvanic reduction of metallic precursor cations[19]. Here we report on the synthesis of bulk nanoporous Al (NPA) as a platform for UV plasmonics. We will show how the porous matrix can support LSPR enabling significant enhancements in both fluorescence and Raman spectroscopy. Finally, numerical simulations will be used to describe the effect.

**2. Materials and Methods**

*Materials:* 7-hydroxil-4-coumarin acetic acid 97%, Aluminum Chloride ($AlCl_3$) anhydrous, powder, 99.99% trace metals basis, 1-Ethyl-3-methylimidazolium chloride 98 % [EMIM]Cl, N-(3-Dimethylaminopropyl)-N′-ethylcarbodiimide hydrochloride BioXtra (EDC hydrochloride), N-Hydroxysuccinimide 98% (NHS), (3-Aminopropyl)triethoxysilane 99% (APTES), Dimethyl sulfoxide anhydrous, ≥99.9% (DMSO). All mentioned chemicals were purchased from Sigma-Aldrich. Acetone, anhydrous (max. 0.01% $H_2O$) ≥99.8% was purchased from VWR chemicals.

*Galvanic Replacement Reaction (GRR)* The preparation of the NPA film started from an $Al_2Mg_3$ alloy prepared by melting high purity Al (>99.9%) and Mg (>99.9%) in a resistance furnace under Ar atmosphere. The NPA was prepared by means of GRR by immersing the $Al_2Mg_3$ sample in an $[EMIM]^+Al_2Cl_7^-$ ionic liquid, which was prepared by mixing $AlCl_3$ and 1-ethyl-3-methylimidazolium chloride, [EMIM]Cl at room temperature with a molar ratio of 2:1. The sample has been kept immersed in the ionic liquid for 8 hours in order to ensure the formation of the bulk porous structure.

*Etching by chemical dealloying (acid treatment)*: Three additional samples of $Al_2Mg_3$ alloy prepared by melting were dipped in a 1 M methanol solution of acetic acid for 30; 60 and 360 min, respectively. The samples were then washed with methanol. Immediately after the acidic treatment, the samples were brought inside the glovebox.

*X-ray diffraction (XRD)* were performed by means of a PANalytical Empyrean X-ray diffractometer equipped with a 1.8kW CuK$\alpha$ ceramic X-ray tube, PIXcel3D 2x2 area detector and operating at 45 kV and 40 mA. The diffraction patterns were collected in air at room temperature using Parallel-Beam (PB) geometry and symmetric reflection mode. XRD data analysis was carried out using HighScore 4.18 software from PANalytical.

*Energy-dispersive spectroscopy (EDS)* was performed within a JEOL JSM-7500LA SEM (JEOL, Tokyo, Japan), equipped with a cold field-emission gun (FEG), operating at 5 kV acceleration voltage. We measured the film composition through an Oxford instrument EDS setup (X-Max, 80 mm²) Energy-dispersive spectroscopy (EDS, Oxford instrument, X-Max, 80 mm²). The measurements were performed at 8 mm working distance, 5 kV acceleration voltage and 15 sweep count for each sample. EDS spectra from three different positions have been collected for each sample. In order to analyze the spectra, we used Aztec 1.2 software®, with automatic calibration of the standards and background subtraction. The instrument was calibrated with a Microanalysis Standard As-02756-AB 59 Metals & Minerals Carousel Serial HM, by SPI. For all the elements we analyzed the K-alpha lines. For each element, the composition percentages measured in different positions differed one from the other by less than 1%,

*Samples functionalization with APTES:* The samples were shacked overnight, at room temperature, in 4% solution of APTES in acetone inside glove box. They were then washed in acetone and dried inside the glove box.

*Dye activation and attachment on the substrate*: 7-hydroxyl-4-coumarin acetic acid (0.01 mmol) was solubilized in DMSO (1 mL) reaching 10 mM as final concentration. EDC (0.02 mmol) was added and stirred at room temperature for 15 minutes. Subsequently, NHS (0.03 mmol) was added and the solution was stirred for additional 45 minutes. After 1 hour, the mixture was diluted in acetone reaching a final concentration of 1 mM of dye. 1 mL of this solution was added to each substrate in a 3 mL glass vial and 1 mL was added in an empty vial as reference for the dye quantification. The vials were shaken at room temperature in glove box overnight. The substrates were carefully withdrawn and washed with acetone. The remaining solution of coumarin was kept for dye quantification: the absorbance at 360 nm was used to determine the amount of coumarin present in each solution using as reference 1 mM solution of activated coumarin. The obtained data were used to estimate the amount of coumarin covalently bound to the substrate surface. Dye quantification was preferentially achieved by measuring dye absorption since this value is not affected by self-quenching, life-time and fluorescence decay which characterize instead fluorescence measurements.

*Fluorescence measurement*: Once washed, all substrates were placed in a plate, a drop of phosphate buffer (PBS) was added on the surface in order to make a measurement under wet conditions, and the fluorescence spectrum was recorded with a 360 nm excitation wavelength ($\lambda_{ex}$). The fluorescence signal was measured selecting specific sections of each well ($\lambda_{ex}:\lambda_{em}$ 360:460 nm) in order to better evaluate the efficiency of the functionalization. A Tecan Infinite M200 instrument was used for recording the fluorescence spectrum and for the fluorescence measurement of the substrates.

*RAMAN*: UV Resonant Raman (UVRR) measurements have been carried out at the Elettra synchrotron radiation facility. A complete description of the experimental apparatus can be found elsewhere.[20] A 266 nm laser source has been employed. The beam reaching the sample was approximately 25 μW. The Raman scattering signal was collected with a backscattering configuration. A Czery-Turner spectrometer with focal length of 750 mm, coupled with an holographic reflection grating of 1800 g/mm and with a Peltier-cooled back-thinned CCD was employed to get the Raman signal. Spectral resolution was set to 8 cm-1. Raman frequencies were calibrated by means of cyclohexane spectra.[21] The samples have been functionalized with commercial salmon sperm DNA (Sigma-Aldrich). It was diluted in pure water up to a concentration of 10 ug/ml (i.e. 33 uM on nucleobases molar concentration), then deposited by drop casting (drop of 5 ul) on a rough Aluminum and on a NPA samples. Considering 1 mm$^2$ of surface covered by the drop it corresponds to a superficial concentration of 20 nucleobases per nm$^2$ in case of homogenous distribution when water is evaporated.

*FEM simulations*: to explore the plasmonic properties of porous Al in the UV range, a numerical investigation of the electromagnetic response of such material has been conducted. A finite element method (FEM) commercial software, COMSOL Multiphysics, has been employed to model the optical behaviour of the nanoporous metal. In particular, following a procedure we recently reported [5], nanometric pores and irregularities of NPA have been numerically reproduced by means of SEM images from experimental samples. Due to their crucial role in generating the highly localized field enhancement, a proper description of the material's features is strictly necessary. Accordingly, the brightness of SEM images of the metal surface cross sections is used to encode the information concerning the profile of a nanoporous film. Upon uploading in the software environment, a SEM image of the metal surface is treated as a two-dimensional function, with values between 0 (where there is no metal) and 1 (where pure metal optical properties are considered), which defines a map of permittivity of the modelled geometry. NPA can be, thus, optically described by the following 2D weighted permittivity function:

$$\varepsilon = \begin{cases} \varepsilon_{metal} * map(x,y) & if \quad map(x,y) > ths \\ \varepsilon_{background} & if \quad map(x,y) < ths \end{cases}$$

where map(x,y) is the spatial function built from the SEM image of the NPA cross section, $\varepsilon_{background}$ the dielectric constant of the considered surrounding environment (here air, therefore equal to 1), $\varepsilon_{metal}$ the dielectric function of the metal (Al for NPA, Au for nanoporous gold (NPG) used a standard for comparison) and ths a threshold value, used to define the numerical boundary between metal and background in the geometry and chosen to maximize the resemblance between model and sample. In the present work, the value ths=0.6 has been set to achieve the most realistic profile. The electromagnetic problem is solved in frequency domain for $\lambda$=260 nm and $\lambda$=350 nm. A port analysis has been employed to simulate an impinging monochromatic plane, propagating in the direction perpendicular to the surface and polarized along the direction parallel to the NPA substrate. Continuity periodic boundary conditions and perfectly matched layers have been defined on the lateral sides and on both the top and bottom of the structure respectively.

## 3. Results and discussion

As previously mentioned, Yang et al.[17] have successfully converted foils of $Al_2Mg_3$ into NPA by exposing them to an ionic liquid. The GRR proceeds as follows:

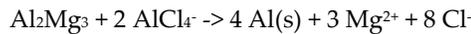

$Al_2Mg_3 + 2\ AlCl_4^- \rightarrow 4\ Al(s) + 3\ Mg^{2+} + 8\ Cl^-$

This system shows very peculiar properties from the standpoint of materials preparation: (i) The GRR causes exhaustive replacement of Mg by Al; (ii) oxide formation and hydrogen evolution are excluded or minimized by operation in ionic liquid; (iii) Al deposition occurs inward, towards the bulk of the foil loosing Mg, and not outwards (towards the electrolyte) as one would normally expect; (iv) The process is a homogeneous plating with fresh Al of the inner nanostructures resulting from Mg loss and Al atoms rearrangement. The resulting product is a bulk, porous, rather pure Al layer. A crucial role for the inward-growth appears to result from two facts: (i) the volume of solid phase decreases during GRR, leaving space for electrolyte permeation of the porous structure and mass transport in the pores; (ii) the metal ion is carried by negatively charged (anionic) species, moving inwards towards the receding reaction front.

*3.1 Structural properties*

Figure 1 shows some examples of NPA morphology obtained from GRR by employing the alloy prepared from melted Al and Mg. The obtained morphology confirms previous results reported in [17]. Comparting this sample with porous structures obtained by means of chemical dealloying[5], nanoporous Al/Mg prepared via dealloying from a co-sputtered film shows pores of dimension down to few tens of nm, which can be modulated acting on the starting composition of the alloy and on the dealloying condition. The samples obtained by means of GRR, on the contrary, are much less uniform in term of morphology and present pores that are an order of magnitude larger. In order to verify that the different morphology was not due to the different procedure of deposition of the alloy (in the previous case we used co-sputtering deposition from Al and Mg targets), chemical dealloying in acetic acid has been here performed on $Al_2Mg_3$ prepared via melting. The results confirm that, while it's possible to achieve nanoporous films with pores down to tens of nm, the chemical dealloying process does not enable the complete etching of Mg and the final oxide contents within the porous samples are always significant.

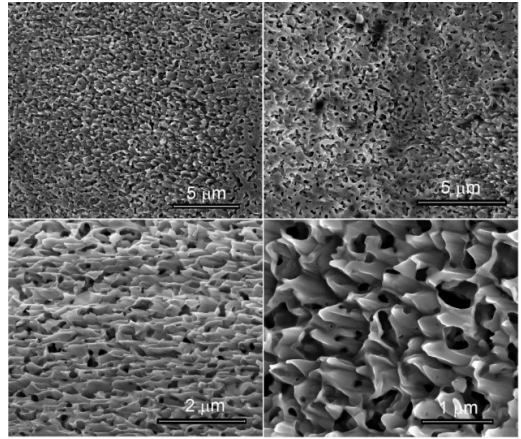

**Figure 1.** SEM micrographs of the GRR sample, displaying different magnifications taken from different area of the same sample.

The compositions, obtained from EDS are reported (the spectra as recorded by the instrument are reported in the SI) in Table 1. The oxygen content around 14% confirms that the obtained material is metallic Al. More importantly, the GRR enables the complete etching of Mg from the starting alloy, as confirmed from the XRD analysis reported in SI. The complete removal of Mg has not been possible with the chemical dealloying[5] and this result demonstrates that the GRR actually get a nanoporous Al structure.

**Table 1.** Samples, initial composition x, composition after the GRR as measured by means of EDS.

| Sample | Pristine composition x ($Mg_xAl_{1-x}$) | (EDS) Etched composition (O, Al, Mg) |
|---|---|---|
| GRR | 0.6 | 14%, 84%, 0% |

The strong air reactivity of both Al and Mg suggests the oxidation of the obtained samples surfaces. An understanding of the film surface (within the first 10 nm) oxygen content can be obtained by means of XPS, as reported in Table 2 (details of the measured spectrum are reported in SI).

**Table 2.** Samples, XPS analyses.

| Sample | $Al_2O_3$ (at%) | Al suboxides (at%) | Metallic Al | MgO |
|---|---|---|---|---|
| GRR | 68.9% | 10.5% | 20.6% | -- |

*3.2 Optical and plasmonic properties*

The first characterization of the NPA optical performances has been done by measuring the reflectance in the spectral range between 200 and 2000 nm. Due to the high roughness of the samples, in order to collect the reflectance spectrum a spectrophotometer equipped with an integrating sphere has been used. In this way the direct and diffuse reflectance can be collected. Fig. 2 reports the spectrum obtained from the described samples in comparison with the spectrum of an $Al_2O_3$ sample. This allows to clearly evaluate the features related to the high porosity and to the oxidation of the sample. The presence of an absorption band at 800 nm, ascribable to the Al interband transition, also confirms the pure metallic structure of the film. Following a procedure recently used for NPG[22], a

Drude – Lorentz model based fit has been performed on the reflectance spectrum in order to obtain an estimation of the dielectric constants of samples (Fig. 2b).

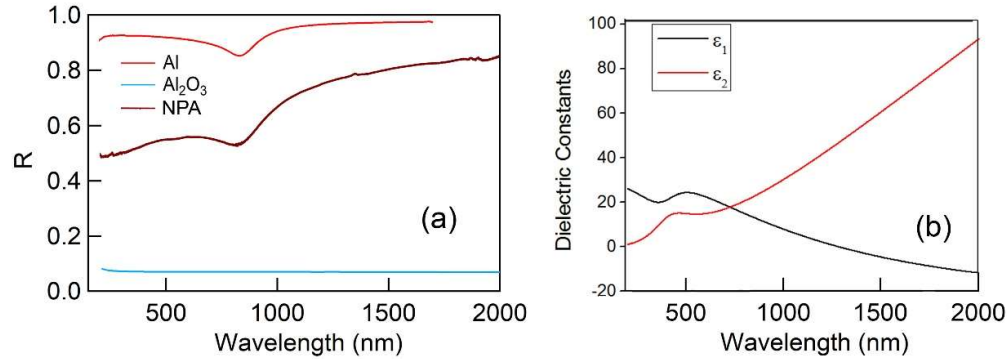

**Figure 2.** (a) Reflectance spectra of Al, Al$_2$O$_3$ and NPA, respectively; (b) Dielectric Constants of NPA obtained from Drude-Lorentz best-fit.

As previously stated[5], this nanoporous samples prepared in metallic Al can find applications in field-enhanced spectroscopies. In particular, enhanced fluorescence (FE) and SERS can be probed in order to evaluate the potential performances of the prepared films.
In order to perform FE experiments, the NPA samples have been functionalized according the protocol reported in Methods. In particular, we chose to use Coumarins as reported dyes. They are well-known dyes widely exploited for their optical properties and for the development of "off-on" switchable fluorescent biosensors[23]. 7-hydroxy-4-coumarin acetic acid was selected for the evaluation of the FE achieved with the here discussed alumina-based porous substrates due to its fluorescent properties. Firstly, this dye absorbs in the UV range (360 nm) and secondly it has a suitable Stokes shift (100nm) which makes easier to measure the fluorescence of the dye on the substrate hence avoiding interferences related to the substrate itself. Furthermore, this dye already proved its applicability for the development of switchable sensors, therefore the possibility of enhancing its fluorescence signal could potentially lead to intriguing improvements in terms of sensitivity and limit of detection (LOD) of such fluorescent-based detection approaches[24]. The substrates was firstly treated with APTES (3-Aminopropyl)triethoxysilane) in order to have amino groups exposed on the surface. These amino groups were subsequently used as anchors to covalently attach the carboxylated-coumarin previously activated using EDC-NHS as coupling agents. A scheme of the different functionalization steps can be found in[5]. The same protocol was performed on NPA samples and on a reference substrate made of rough Al. This allows to compare the FE achievable using our substrates and a well-known film with good properties in the UV spectral range. The comparison was accomplished considering both the fluorescent signal measured for each kind of substrate and the corresponding amount of dye effectively attached on the surface. In particular, we evaluated the enhancing efficiency by means of the FE factor calculated by dividing the fluorescent signal by the calculated amount of coumarin present at the surface. The FE factor is normalized to the value determined for rough Al. In order to be able to compare the calculated concentration of coumarin attached on the surface and the related fluorescence, the relative values for both dye's concentration and fluorescence signal were considered. As noticeable in Figure 3A, the fluorescent signal was almost 5 times higher than the concentration of dye for GRR, suggesting that this substrate gives a significant FE with respect to rough Al. The FE factor shown in Figure 3A better defines the relationship between concentration of dye and fluorescent signal, allowing to determine the fluorescence enhancement for each substrate here considered.

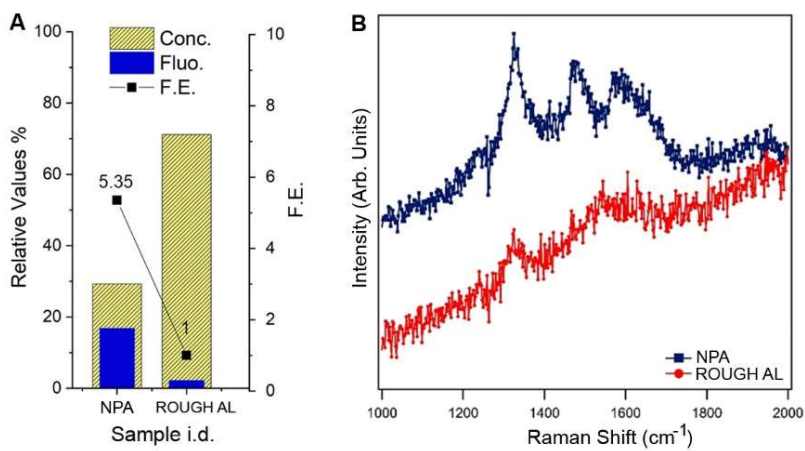

**Figure 3.** A) Experimental FE in GRR NPA compared to standard rough Al sample; B) Experimental UVRR spectra of salmon sperm DNA deposited by drop casting on rough Al (red curve) and NPA (blue curve) substrates.

The potential application of the prepared NPA substrate as SERS platform has been evaluated by measuring the UV Raman spectrum from commercial genomic DNA. The concentration used for the functionalization of NPA and rough Al reference samples correspond to about 20 nucleotides per $nm^2$. Figure 3B shows the UVRR spectrum obtained by salmon sperm DNA deposited on reference Al and on NPA. Each spectrum corresponds to an average of 8 spectra collected in different positions of the drop, equally distributed from the center to the border of the drop limit. Both the spectra evidences the typical DNA vibrational features. More specifically the peaks at 1337 $cm^{-1}$ is addressed to adenine, the ones at 1484 $cm^{-1}$ and 1580 $cm^{-1}$ to both adenine and guanine, while the bump at 1655 $cm^{-1}$ is mainly a thymine contribution[25]. The NPA peaks results enhanced by a factor at least of 5 with respect to those of the spectrum acquired on reference Al surface.

To be noted, the obtained enhancement in FE and in SERS are even more significant if we consider that the rough Al samples used as reference is known to enhance the fluorescence by a factor between 2 and 10[26]. Finally rough Al has been also verified to provide enhancement in Raman in the UV spectral region[27].

*3.3 Numerical simulations*

Finally, the optical behaviour of NPA in the UV range has been investigated numerically. The field confinement induced by the nanoporous structure of the film is determined by means of a 2D electromagnetic computation (see Methods for details), and its spatial distribution is shown in Fig. 4 across a sample section, for different values of excitation wavelength. To highlight the better performances achieved in the considered spectral range by NPA if compared to standard plasmonic materials, e.g. Au, the same computation has been performed on NPG. In doing so and similarly to how we recently reported [5], the quantity $|E/E_0|$ is computed across the same geometry, namely generated by using the same SEM image of the experimental sample, but optically describe by Al (Figs. 4A and 4C) and Au (Figs. 4B and 4D) dielectric function respectively. To further stress the higher localisation of the field in the case of NPA, scales for field enhancement of Fig.4 are also the same. This choice in displaying data makes the gain in using NPA instead of NPG even more evident. Indeed, this shows that, for a given nanostructure of pores and spatial features of a surface, a much higher confinement can be induced in the considered spectral region when Al is used.

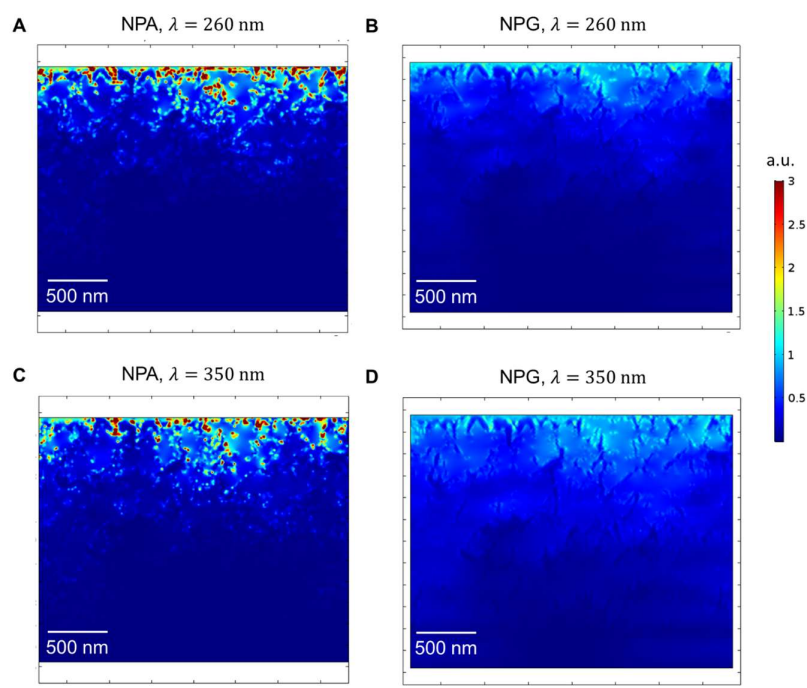

**Figure 4.** Comparison of performances between nanoporous Al and Au in the UV range. Here, the numerically-computed field enhancement in NPA (A,C) and NPG (B,D) films, at excitation wavelengths $\lambda = 260$ nm (A,B) and $\lambda = 350$ nm (C,D).

## 4. Conclusions

In conclusion, here we reported on GRR based procedures to prepare NPA films from an alloy of Al and Mg. Although present, the oxide layer is stable in time and under it a porous structure of metallic Al is always present. Optical spectroscopies have been used to evaluate the dielectric constants of the prepared NPA and to verify its potential applications as UV plasmonic material. In particular we verified that NPA films can be used as an interesting platform both in UV fluorescence and UV SERS. Enhancement factors about 5 with respect to a standard rough Al sample have been observed in both the cases.

**Author Contributions:** D.G., G.G., S.C. and P.P. fabricated and characterized the structures; F.D. and M.P. helped in the optical characterization; R.P.Z., A.S., and A.A. performed the e.m. simulations; W.Y., H.-J.J., R.K., helped in manuscript preparation. D.G. supervised the work.

**Conflicts of Interest:** The authors declare no conflict of interest.

## References

1. Stockman, M.I.; Kneipp, K.; Bozhevolnyi, S.I.; Saha, S.; Dutta, A.; Ndukaife, J.; Kinsey, N.; Reddy, H.; Guler, U.; Shalaev, V.M.; et al. Roadmap on plasmonics. *J. Opt. (United Kingdom)* **2018**, *20*.

2. Gerard, D.; Gray, S.K. Aluminium plasmonics. *J. Phys. D. Appl. Phys.* **2015**, *48*.

3. McMahon, J.M.; Gray, S.K.; Schatz, G.C. Ultraviolet Plasmonics: The Poor Metals Al, Ga, In, Sn, Tl, Pb, and Bi. **2009**, 1–33.

4. Ding, T.; Sigle, D.O.; Herrmann, L.O.; Wolverson, D.; Baumberg, J.J. Nanoimprint lithography of Al nanovoids for deep-UV SERS. *ACS Appl. Mater. Interfaces* **2014**, *6*, 17358–17363.

5. Ponzellini, P.; Giovannini, G.; Cattarin, S.; Zaccaria, R.P.; Marras, S.; Prato, M.; Schirato, A.; D'Amico, F.; Calandrini, E.; De Angelis, F.; et al. Metallic Nanoporous Aluminum–Magnesium Alloy for UV-Enhanced Spectroscopy. *J. Phys. Chem. C* **2019**, *123*, 20287–20296.

6. Teale, F.W.J.; Weber, G. Ultraviolet fluorescence of the aromatic amino acids. *Biochem. J.* **1957**, *65*, 476–482.

7. Fothergill, S.M.; Joyce, C.; Xie, F. Metal enhanced fluorescence biosensing: From ultra-violet towards second near-infrared window. *Nanoscale* **2018**, *10*, 20914–20929.

8. Gutiérrez, Yael; Alcaraz de la Osa, Rodrigo; Ortiz, Dolores ; Saiz, José María; González, Francisco; Moreno, F. Plasmonics in the Ultraviolet with Aluminum, Gallium, Magnesium and Rhodium. *Appl. Sci.* **2018**, *8*, 64.

9. Everitt, H.O.; Gutierrez, Y.; Sanz, J.M.; Saiz, J.M.; Moreno, F.; Gonzalez, F.; Ortiz, D. How an oxide shell affects the ultraviolet plasmonic behavior of Ga, Mg, and Al nanostructures. *Opt. Express* **2016**, *24*, 20621.

10. Ahmadivand, A.; Sinha, R.; Kaya, S.; Pala, N. Rhodium Plasmonics for Deep-Ultraviolet Bio-Chemical Sensing. *Plasmonics* **2016**, *11*, 839–849.

11. Appusamy, K.; Jiao, X.; Blair, S.; Nahata, A.; Guruswamy, S. Mg thin films with Al seed layers for UV plasmonics. *J. Phys. D. Appl. Phys.* **2015**, *48*.

12. Watanabe, K.; Saito, Y.; Honda, M.; Kumamoto, Y.; Kawata, S.; Taguchi, A. Indium for Deep-Ultraviolet Surface-Enhanced Resonance Raman Scattering. *ACS Photonics* **2014**, *1*, 598–603.

13. Knight, M.W.; King, N.S.; Liu, L.; Everitt, H.O.; Nordlander, P.; Halas, N.J. Aluminum for plasmonics. *ACS Nano* **2014**, *8*, 834–840.

14. McClain, M.J.; Zhou, L.; Tian, S.; Neumann, O.; Nordlander, P.; Halas, N.J.; Zhang, C.; Yang, X. Aluminum Nanocrystals: A Sustainable Substrate for Quantitative SERS-Based DNA Detection. *Nano Lett.* **2017**, *17*, 5071–5077.

15. Martin, J.; Plain, J. Fabrication of aluminium nanostructures for plasmonics. *J. Phys. D. Appl. Phys.* **2015**, *48*.

16. Ruffato, G.; Garoli, D.; Cattarin, S.; Barison, S.; Natali, M.; Canton, P.; Benedetti, A. Microporous and Mesoporous Materials Patterned nanoporous-gold thin layers : Structure control and tailoring of plasmonic properties. *MICROPOROUS MESOPOROUS Mater.* **2012**, *163*, 153–159.

17. Yang, W.; Zheng, X.-G.; Wang, S.-G.; Jin, H.-J. Nanoporous Aluminum by Galvanic Replacement: Dealloying and Inward-Growth Plating. *J. Electrochem. Soc.* **2018**, *165*, C492–C496.

18. Corsi, J.S.; Fu, J.; Wang, Z.; Lee, T.; Ng, A.K.; Detsi, E. Hierarchical Bulk Nanoporous Aluminum for On-Site Generation of Hydrogen by Hydrolysis in Pure Water and Combustion of Solid Fuels. *ACS Sustain. Chem. Eng.* **2019**, *7*, 11194–11204.

19. Sun, Y.; Mayers, B.T.; Xia, Y. Template-Engaged Replacement Reaction: A One-Step Approach to the Large-Scale Synthesis of Metal Nanostructures with Hollow Interiors. *Nano Lett.* **2002**, *2*, 481–485.

20. D'Amico, F.; Saito, M.; Bencivenga, F.; Marsi, M.; Gessini, A.; Camisasca, G.; Principi, E.; Cucini, R.; Di Fonzo, S.; Battistoni, A.; et al. UV resonant Raman scattering facility at Elettra. *Nucl. Instruments Methods Phys. Res. Sect. A Accel. Spectrometers, Detect. Assoc. Equip.* **2013**, *703*, 33–37.

21. McCreery, R.L. Photometric Standards for Raman Spectroscopy. In *Handbook of Vibrational Spectroscopy*; Griffiths, P.R., Ed.; John Wiley & Sons, Ltd: Chichester, UK, 2006.

22. Garoli, D.; Calandrini, E.; Bozzola, A.; Toma, A.; Cattarin, S.; Ortolani, M.; De Angelis, F. Fractal-Like Plasmonic Metamaterial with a Tailorable Plasma Frequency in the near-Infrared. *ACS Photonics* **2018**, *5*, 3408–3414.

23. Dai, X.; Wu, Q.-H.; Wang, P.-C.; Tian, J.; Xu, Y.; Wang, S.-Q.; Miao, J.-Y.; Zhao, B.-X. A simple and effective coumarin-based fluorescent probe for cysteine. *Biosens. Bioelectron.* **2014**, *59*, 35–39.

24. Giovannini, G.; Hall, A.J.; Gubala, V. Coumarin-based, switchable fluorescent substrates for enzymatic bacterial detection. *Talanta* **2018**, *188*, 448–453.

25. Zucchiatti, P.; Latella, K.; Birarda, G.; Vaccari, L.; Rossi, B.; Gessini, A.; Masciovecchio, C.; D'Amico, F. The quality is in the eye of the beholder: The perspective of FTIR and UV resonant Raman spectroscopies on extracted nucleic acids. *J. Raman Spectrosc.* **2018**, *49*, 1056–1065.

26. Ray, K.; Chowdhury, M.H.; Lakowicz, J.R. Aluminum nanostructured films as substrates for enhanced fluorescence in the ultraviolet-blue spectral region. *Anal. Chem.* **2007**, *79*, 6480–6487.

27. Dörfer, T.; Schmitt, M.; Popp, J. Deep-UV surface-enhanced Raman scattering. *J. Raman Spectrosc.* **2007**, *38*, 1379–1382.